\documentclass[proceedings]{JHEP37}
\usepackage{eurosym}
\usepackage{amssymb}
\usepackage{amsfonts,cite}
\usepackage{amsmath}
\usepackage{graphicx}
\usepackage{multirow}
\usepackage{epsfig}

\newbox\mybox

\newcommand\fverb{\setbox\mybox=\hbox\bgroup\verb}
\newcommand\fverbdo{\egroup\medskip\noindent\fbox{\unhbox\mybox}\ }
\newcommand\fverbit{\egroup\item[\fbox{\unhbox\mybox}]}
\conference{Real energies and Berry phases in all PT-regimes in non-Hermitian theories}
\abstract{We demonstrate that the existence of a Hermitian time-dependent intertwining operator that maps the non-Hermitian time-dependent energy operator to its Hermitian conjugate and its right to its left eigenstates guarantees the reality of the instantaneous energies. This property holds throughout all three $\cal{PT}$-regimes, in the time-independent scenario referred to as the $\cal{PT}$-symmetric regime, the exceptional point and the spontaneously broken $\cal{PT}$-regime. We also propose a modified adiabatic approximation consisting of an expansion of the wavefunctions in terms the instantaneous eigenstates of the energy operator, instead of the usually used eigenfunctions of the Hamiltonian. We show that this proposal always leads to real Berry phases. We illustrate the working of our general proposals with two explicit examples for a time-dependent non-Hermitian spin model.}  

\title{Real energies and Berry phases in all PT-regimes in time-dependent non-Hermitian theories}
\author{Andreas Fring$^\bullet$, Takanobu Taira$^\circ$ and Rebecca Tenney$^\bullet$\\
 $\bullet$ Department of Mathematics, City, University of London, Northampton Square,\\ $\,\,$ London EC1V 0HB, UK \\
 $\circ$ Research Fellow of Japan Society for Promotion of Science, Institute of Industrial \\ $\,\,$ Science, The University of Tokyo
 5-1-5 Kashiwanoha,
  Kashiwa 277-8574, Japan\\
 
E-mail: a.fring@city.ac.uk, taira904@iis.u-tokyo.ac.jp, rebecca.tenney@city.ac.uk}

\input{tcilatex}
\begin{document}

\section{Introduction}
Time-independent non-Hermitian $\cal{PT}$-symmetric/quasi-Hermitian quantum systems are characterised by three different regimes in their parameter space in which they exhibit fundamentally distinct behaviour. In their $\cal{PT}$-symmetric regime the eigenspectra of the Hamiltonian are all real, in the spontaneously broken $\cal{PT}$-regime at least two of the eigenvalues occur in complex conjugate pairs and at the transition point between the two regimes, the exceptional point, the eigenvalues and eigenstates coalesce. Thus this point is different from standard degeneracy where only the eigenvalues coalesce. It is well-known that the explanation for this behaviour can be attributed to the existence of an antilinear symmetry operator \cite{EW} of which $\cal{PT}$ \cite{Bender:1998ke}, a simultaneous reflection in space and time is an example. In the symmetric regime the $\cal{PT}$-operator commutes with the Hamiltonian and their eigenstates are identical up to a phase, whereas in the broken regime the latter property no longer holds.  

For systems described by explicitly time-dependent Hamiltonians it was recently observed \cite{AndTom3,fring2019eternal,fring2020time,frith2020exotic,fring2021perturb,alsalam2022,huang2022solvable} that they are no longer separated by exceptional points, even though the qualitative behaviour in these regimes in parameter space might still differ. Especially remarkable is the observation that the instantaneous energy eigenvalues become real in the spontaneously broken regime. Other physical quantities, such as the von Neumann entropy, inherit this behaviour and lead to new physical effects when crossing from one regime to the other \cite{fring2019eternal,frith2020exotic}. Up to now these features were only observed in case-by-case studies of specific models, but a generic explanation for such a behaviour was still missing. The main purpose of this manuscript is to provide such an explanation.

Central to the proper treatment of explicitly time-dependent non-Hermitian Hamiltonian systems is a clear distinction between the energy operator and the Hamiltonian. Unlike as in the Hermitian case they are no longer identical with the latter not even being an observable quantity \cite{CA,time1,time6,time7,fringmo,mostafazadeh2020time,BeckyAnd2,fring2021perturb,fring2021exactly,fring2021infinit,most2018en}, see also \cite{fring2022intro} and additional references therein. Here we build further on this difference and propose a new type of adiabatic approximation that will always lead to real Berry phases for non-Hermitian $\cal{PT}$-symmetric/quasi-Hermitian quantum systems.

Our manuscript is organised as follows: In section 2 we provide the general argument that the existence of a time-dependent intertwining operator with certain properties guarantees the reality of the instantaneous energy eigenstates. In section 3 we propose an alternative adiabatic expansion in terms of the eigenvectors of the energy operator that will always lead to real geometric phases when the inner product is appropriately taken with a new metric operator. In section 4 we present a worked out example for a spin model in terms of Pauli matrices with complex time-dependent coefficients for two different types of metric operators. Our conclusions are stated in section 5.

\section{Real instantaneous energy eigenvalues in time-dependent non-Hermitian systems}
We start by noticing that the time-dependent quasi-Hermiticity relation for a non-Hermitian Hamiltonian $H(t)$ can be re-written formally as the standard quasi-Hermiticity relation for the instantaneous energy operator $\tilde{H}(t)$
\begin{equation}
	i\hbar \partial_t \rho = H^\dagger \rho - \rho H, \qquad  \Leftrightarrow \qquad  \tilde{H}^\dagger \rho = \rho \tilde{H}, \quad \text{with} \,\,\, \tilde{H} := H + i \hbar \eta^{-1} \partial_t \eta.  \label{TDQHE}
\end{equation}
Here $\rho(t)=\eta^\dagger(t) \eta(t)$ is a time-dependent positive definite metric and $\eta$ denotes the time-dependent Dyson map, see recent review \cite{fring2022intro} and references therein for more details on these quantities. When given the Hamiltonian $H(t)$ as a starting point, one needs to stress that the left relation in (\ref{TDQHE}) has to be solved first for $\rho(t)$ before the expression for the energy operator $\tilde{H}$ can actually be written down. Thus, unless very strong assumptions on the Dyson map are made, the operator $\tilde{H}$ is a meaningless starting point. There might also be solutions to the right version in (\ref{TDQHE}) that are not positive definite, which we denote as $\tilde{{\cal P}}(t)$ in loose analogy to the time-independent version where the corresponding operator can be identified as the parity operator. However, we do not require this operator to be an involution, thus in general we have $\tilde{{\cal P}}^2(t) \neq \mathbb{I} $. Notice further that due to the fact that $\tilde{{\cal P}}(t)$ is negative definite, $\tilde{{\cal P}}(t)<0$, it does not factorise into Dyson maps $\tilde{{\cal P}}(t) \neq \eta^\dagger \eta $, so that the left relation in (\ref{TDQHE}) does not hold when $\rho$ is replaced with $\tilde{{\cal P}}(t)$. 

We can now define two types of time-dependent ${\cal C}(t)$-operators, which has been introduced as the operator that relates negative definite to positive definite solutions of the quasi-Hermiticity relation in the time-independent \cite{Bender:2002vv} as well as in the time-dependent scenario \cite{fring2022time}. The first option for the latter case was introduced in \cite{fring2022time}, where ${\cal C}(t)$ was expanded in terms of the right and left eigenstates,  $\vert \hat{\psi}(t) \rangle$ and $\vert \hat{\phi}(t) \rangle$, of the non-Hermitian Hamiltonian $H(t)$ 
\begin{equation}
	H(t) \vert \hat{\psi}(t) \rangle = E \vert \hat{\psi}(t) \rangle, \qquad 
	H^\dagger	\vert \hat{\phi}(t) \rangle = E \vert \hat{\phi}(t) \rangle, \qquad  \langle \hat{\phi}(t) \vert \hat{\psi}(t) \rangle = \mathbb{I},
\end{equation}
as
\begin{equation}
	{\cal C}(t) := \sum_n \hat{s}_n \vert \hat{\psi}_n(t) \rangle \langle  \hat{\phi}_n(t) \vert
	= \sum_n \hat{s}_n \vert \hat{\phi}_n(t) \rangle \langle  \hat{\psi}_n(t) \vert
	={\cal P} \hat{\rho}(t) =\hat{\rho}^{-1}(t) {\cal P} . \label{defc}
\end{equation}
Here $\hat{s}_n = \pm 1$ are the signatures of this expansion and ${\cal P}$ is the time-independent parity operator with ${\cal P}^{2} = \mathbb{I}$. The operator ${\cal{ C}}$ is involutory but, unlike as in the time-independent case, does not commute with the time-dependent Hamiltonian
\begin{equation}
	{\cal{ C}}^2(t) = \mathbb{I}, \qquad \text{and} \qquad i \hbar \partial_t {\cal{ C}}(t)=[  H(t),{\cal{ C}}(t)]. \label{cprop}
\end{equation}
Besides relating a time-independent negative definite solution ${\cal P}$ to the pseudo-Hermiticity relation to a positive time-dependent definite metric operator $\hat{\rho}(t)$, due to the first relation in (\ref{cprop}) this ${\cal C}$-operator can formally also be identified with a Lewis-Riesenfeld invariant as shown in \cite{fring2022time} when we identify $\hat{\rho}(t)$ with $\rho(t)$.  

  Alternatively, we can define a different operator in terms of the right and left eigenstates, $\vert \tilde{\psi}(t) \rangle$ and $\vert \tilde{\phi}(t) \rangle$, of the energy operator $\tilde{H}$ defined in (\ref{TDQHE})
\begin{equation}
	\tilde{H} \vert \tilde{\psi}(t) \rangle = \tilde{E} \vert \tilde{\psi}(t) \rangle, \qquad 
  	\tilde{H}^\dagger	\vert \tilde{\phi}(t) \rangle = \tilde{E} \vert \tilde{\phi}(t) \rangle, \qquad  \langle \tilde{\phi}(t) \vert \tilde{\psi}(t) \rangle = \mathbb{I}, \label{evequn}
\end{equation}
that relates the non-positive definite time-dependent to the positive definite time-dependent solutions as
\begin{equation}
      \tilde{{\cal C}}(t) := \sum_n \tilde{s}_n \vert \tilde{\psi}_n(t) \rangle \langle  \tilde{\phi}_n(t) \vert
      = \sum_n \tilde{s}_n \vert \tilde{\phi}_n(t) \rangle \langle  \tilde{\psi}_n(t) \vert
      =\tilde{{\cal P}}^{-1}(t)\rho(t)=\rho^{-1}(t)\tilde{{\cal P}}(t) . \label{defct}
\end{equation}
Here $\tilde{s}_n = \pm 1$ are the signatures of this expansion. The two expansions (\ref{defc}) and  (\ref{defct}) are obviously different. In the time-independent version this ambiguity does not exist as the two operators $H$ and $\tilde{H}$, and therefore its eigenstates coincide. It is easily verified that $\tilde{{\cal C}}(t)$, as defined in (\ref{defct}), is an involution operator that commutes with the energy operator,
\begin{equation}
\tilde{{\cal C}}^2(t) = \mathbb{I}, \qquad \text{and} \qquad [ \tilde{{\cal C}}(t), \tilde{H}(t)]=0. \label{cprop}
	\end{equation}
We now make the following observation: {\em When the time-dependent non-Hermitian energy operator $\tilde{H}(t)$ is quasi-Hermitian with regard to the action of the time-dependent Hermitian intertwining operator $\tilde{{\cal P}}$  that relates its right eigenvectors to its left eigenvectors, then the instantaneous energies $\tilde{E}(t)$ are real. Thus, if the properties}
\begin{equation}
  i) \,\,\tilde{{\cal P}}  \tilde{H} =  \tilde{H}^\dagger \tilde{{\cal P}} , \qquad    ii) \,\,  \tilde{{\cal P}} \vert \tilde{\psi} \rangle    = \alpha \vert \tilde{\phi} \rangle ,  \qquad   iii) \,\,  \tilde{{\cal P}}= \tilde{{\cal P}}^\dagger, \,\,  \,\,
    \alpha \in \mathbb{R}, \label{PTrel}
\end{equation}
{\em hold, then  $\tilde{E}(t) \in \mathbb{R}$ at any time $t$.}

We easily prove this statement using these three properties together with the standard properties of left and right eigenvectors (\ref{evequn}). We have
\begin{equation}
	\tilde{E}\!\!\overset{\text{(\ref{evequn})}}{=} \!\!
\langle \tilde{\phi} \vert \tilde{H} \vert \tilde{\psi} \rangle
\overset{ii,iii}{=}
\frac{1}{\alpha} \langle \tilde{\psi} \vert  \tilde{{\cal P}} \tilde{H} \vert \tilde{\psi} \rangle 
\overset{i}{=} 
\frac{1}{\alpha} \langle \tilde{\psi} \vert \tilde{H}^\dagger\tilde{{\cal P}}  \vert \tilde{\psi} \rangle
\overset{\text{(\ref{evequn})}}{=} 
\frac{E^*}{\alpha} \langle \tilde{\psi} \vert \tilde{{\cal P}} \vert \tilde{\psi} \rangle  
\overset{ii,iii}{=}\tilde{E}^* \langle \tilde{\phi} \vert \tilde{\psi} \rangle \overset{\text{(\ref{evequn})}}{=}
\tilde{E}^* .
\end{equation}
We notice that we no longer require the use of the time-reversal operator and moreover that property $ii)$ in (\ref{PTrel}) is slightly different from the usual requirement used in the time-independent case where ${\cal{PT}}$ maps right/left eigenvector to right/left eigenvectors up to a phase, see e.g. section 2.4.1 in \cite{fring2022intro}. Thus, in that case $\cal{PT}$ 
is an involution operator, which is no longer the case for the  $\tilde{{\cal P}}$-operator, i.e. $\tilde{{\cal P}}^2 \neq \mathbb{I}$.  This is also the reason why we do not refer to  $\tilde{{\cal P}}$ as a time-dependent parity operator. As we stated, the time-dependent quasi-Hermiticity relation holds in general for the energy operator $\tilde{H}$ whereas the time-independent ${\cal PT}$-symmetry holds for the Hamiltonian $H$, which explains why the spontaneously broken ${\cal PT}$-regime can be mended in the time-dependent scenario. This feature was previously observed for an number of systems
 \cite{AndTom3,fring2019eternal,fring2020time,frith2020exotic,fring2021perturb,alsalam2022,huang2022solvable}, but left unexplained up to now.

\section{Real Berry phases in non-Hermitian systems}
We briefly recall how the geometrical (Berry) phase \cite{berry1984quantal} originates in standard Hermitian systems in order to set the scene for a non-Hermitian generalisation. One considers a quantum mechanical system described by an explicitly time-dependent Hermitian Hamiltonian $h[q(t)]$, where the time-dependence is acquired through a set of parameters $q(t)=(q_1(t),q_2(t),\ldots,q_n(t))$ with period $T$, i.e. $q(0)=q(T)$. The time-dependent Schr\"odinger equation (TDSE) then governs the wave function $\left\vert \chi(t) \right\rangle$ as
\begin{equation}
	h[q(t)] \left\vert \chi(t) \right\rangle = i \hbar \partial_t \left\vert \chi(t) \right\rangle .  \label{TDSEh}
\end{equation}
Crucially, the system is assumed to evolve adiabatically so that at each moment in time one can expand the system in terms of instantaneous orthonormal energy eigenstates as  
\begin{equation}
	h[q(t)] \vert \tilde{\chi}_n(t) \rangle = \tilde{E}_n[q(t)] \vert \tilde{\chi}_n(t) \rangle  . \label{12inst}
\end{equation}
The states $\left\vert \chi(t) \right\rangle$ are then assumed to be expanded as
 \begin{equation}
 	\left\vert \chi(t) \right\rangle = \sum_n c_n(0) e^{i \gamma_n(t)}  e^{i \alpha_n(t)} \vert \tilde{\chi}_n(t) \rangle,  \quad 
 	\alpha_n(t)= -\frac{1}{\hbar} \int_0^t \tilde{E}_n[q(t)], \quad c_n(0)=const \in \mathbb{C}, \label{expand}
 \end{equation}
where one distinguishes between the time-dependent dynamical phases $\alpha_n(t)$ and the geometrical phases $\gamma_n(t)$ at level $n$. Substituting the expansion of $\left\vert \chi(t) \right\rangle$ from (\ref{expand}) into the TDSE (\ref{TDSEh}), together with the orthonormality relation of the instantaneous energy eigenstates $\langle  \tilde{\chi}_n(t) \vert \tilde{\chi}_m(t) \rangle = \delta_{n,m} $, one derives that the geometrical phase has to satisfy 
\begin{equation}
   \dot{\gamma}_n(t) = i \langle \tilde{\chi}_n [q(t)] \vert  \partial_t \tilde{\chi}_n [q(t)] \rangle . \label{12}
\end{equation}
The phase becomes a physical observable when the ray $\vert \tilde{\chi}_n [q(t)] \rangle$ returns to its initial state, i.e. when $\vert \tilde{\chi}_n [q(0)] \rangle=\vert \tilde{\chi}_n [q(T)]\rangle $, one may have picked up a non-vanishing phase difference 
\begin{equation}
	\gamma_n = i \int_0^T \left\langle \tilde{\chi}_n [q(t)] \right\vert \left. \partial_t \tilde{\chi}_n [q(t)] \right\rangle dt= i \oint_ C
	\left\langle \tilde{\chi}_n [q(t)] \right\vert \left. \nabla_q   \tilde{\chi}_n [q(t)] \right\rangle dq, \label{Berryph}
\end{equation}
where $C$ is a closed path traced out in parameter space. It is the latter expression which makes it clear why $\gamma_n$ is referred to as a geometrical phase. In a Hermitian system $\gamma_n$ is known to be always real.

Let us now explain how the above extends to a real geometric phase for pseudo-Hermitian systems.  For this purpose we consider now a non-Hermitian Hamiltonian $	H[q(t)]$ depending on the same time-dependent parameter set $q(t)$ as the Hermitian Hamiltonian $h[q(t)]$ satisfying the TDSE
\begin{equation}
	H[q(t)] \left\vert \psi(t) \right\rangle = i \hbar \partial_t \left\vert \psi(t) \right\rangle ,  \label{TDSEH}
\end{equation}
and the time-dependent Dyson equation
\begin{equation}
	h[q(t)] = \eta(t) 	H[q(t)]  \eta^{-1}(t) + i \hbar \partial_t \eta(t) \eta^{-1}(t), \qquad \text{with} \,\,\, \vert \chi(t) \rangle = \eta(t) \vert \psi(t) \rangle . \label{TDDE}
\end{equation}
Thus, we are not considering here open non-Hermitian systems as for instance in \cite{garrison1988complex,liang2013topological,pap2022unified}. Just as in the Hermitian case we assume that the system evolves adiabatically at each moment in time. However, unlike as for the Hermitian case we have now formally two options to implement the adiabatic assumption, we may either expand $\vert \psi(t) \rangle$ in terms of the instantaneous eigenstates of the Hamiltonian $H$ or the energy operator $\tilde{H}$. This is the same ambiguity already encountered previously for the expansion of the $\cal{C}$ and $\tilde{\cal{C}}$-operators. In \cite{garrison1988complex,liang2013topological,time7} the former option was chosen leading to complex Berry phases, here we chose the latter obtaining always real phases. In  \cite{amaouche2022non} the authors employ the eigenstates of the Lewis-Riesenfeld invariant, obtaining real Berry phases.  In \cite{zhang2019time} the authors showed that one may also obtain real phases by imposing the constraint $H=-i \hbar \eta^{-1} \dot{\eta}$, which is, however, an unnecessary limitation for our definition of the Berry phase. Of course one may also consider non-adiabatic phases and avoid the above expansions \cite{maamache2015periodic,cheniti2020adiabatic,most2018en}. Here we use 
\begin{equation}
	\tilde{H}[q(t)] \vert \tilde{\psi}_n(t) \rangle = \tilde{E}_n[q(t)] \vert \tilde{\psi}_n(t) \rangle  ,  \qquad \text{with} \,\,\, \tilde{H}[q(t)] = H[q(t)] + i \hbar \eta^{-1} \partial_t \eta,
\end{equation}
where $\tilde{H}[q(t)]$ is the energy operator already encountered in (\ref{TDQHE}), rather than the Hamiltonian $H[q(t)]$. The second relation in (\ref{TDDE}) together with the orthonormality of the states $\vert \tilde{\phi}_n(t) \rangle$ implies that the states $\vert \tilde{\psi}_n(t) \rangle$ are orthonormal with regard to the new time-dependent metric $\rho(t) = \eta^\dagger(t) \eta(t)$, that is $ \langle \tilde{\psi}_n(t) \vert \rho(t) \tilde{\psi}_m(t)  \rangle = \delta_{n,m}$. The solutions of the TDSE $\vert \psi(t) \rangle$ then expanded as 
\begin{equation}
	\left\vert \psi(t) \right\rangle = \sum_n c_n(0) e^{i \gamma_n(t)}  e^{i \alpha_n(t)}\vert \tilde{\psi}_n(t) \rangle,   \label{expandpsi}
\end{equation}    
with both phases $\alpha_n$ and $\gamma_n$ being identical to those in the Hermitian case.  

Substituting the expansion (\ref{expandpsi}) into the TDSE (\ref{TDSEH}), together with the orthonormality relation for $ \vert \tilde{\phi}_n(t) \rangle $, we derive that the geometrical phase has to satisfies 
\begin{equation}
	\dot{\gamma}_n(t) = i \left\langle \tilde{\psi}_n [q(t)] \right\vert  \rho(t) \left( \partial_t +  \eta^{-1} \partial_t \eta      \right)        \left\vert  \tilde{\psi}_n [q(t)] \right\rangle . \label{BerrynH1}
\end{equation}
Relation (\ref{BerrynH1}) also follows directly when using $\vert \tilde{\chi}(t) \rangle = \eta(t) \vert \tilde{\psi} (t) \rangle$ in (\ref{12}). Thus we have two alternative ways to compute the geometrical phase $\gamma_i(t)$, i.e. either in terms of quantities related to the non-Hermitian system (\ref{BerrynH1}) or in terms of the quantities of the equivalent Hermitian system (\ref{Berryph}). As both expressions can be converted into each other and the phase (\ref{Berryph}) can be shown to be always real, this implies that the phase computed for the non-Hermitian system in (\ref{BerrynH1}) must also be real. Regarding the physical interpretation associated to the two expansions one should note that the expansion in terms of the eigenvalues of the energy operator provides a clear picture identical to the interpretation in the Hermitian case. However, it is unclear what the commonly used expansion in terms of the eigenstates of the Hamiltonian should be as the Hamiltonian is not even an observable operator and has no chance of becoming one as it is not quasi-Hermitian with regard to a time-dependent metric.

\section{An example: an explicitly time-dependent two level system}

As a sample system we consider the explicitly time-dependent non-Hermitian Hamiltonian of the most general $(2 \times 2)$-matrix form describing a spin system
\begin{equation}
	H(t)=-\frac{1}{2}\left[ \omega \mathbb{I}+ \alpha(t) \sigma _{x}+  \mu(t) \sigma _{y} + \tau(t) \sigma _{z} \right], \qquad\quad \text{with} \,\,\,  \alpha, \mu, \tau \in \mathbb{C}, \label{Hfinitet}
\end{equation} 
with $\sigma_x,\sigma_y, \sigma_z$ denoting standard Pauli matrices. Treating this Hamiltonian initially as time-independent, i.e. we take the parameter $\alpha, \mu, \tau$ for a fixed time, we may identify the parity operator as the non-positive definite solution of the pseudo-Hermiticity equation ${\cal P} H = H^\dagger {\cal P}$ with ${\cal P}^2=\mathbb{I}$. Separating real and imaginary parts as $x=x_r + i x_i$ for $x=\alpha, \mu, \tau$, $x_r,x_i \in \mathbb{R}$ and imposing the constraints 
\begin{equation}
	\alpha_r  \alpha_i =  -\mu_r  \mu_i, \qquad \text{and} \qquad \tau_r =0, \label{const1}
\end{equation} 
we find the solution
\begin{equation}
	{\cal P}= \left(
	\begin{array}{cc}
		0 & 	\frac{\alpha _r- i\mu_r}{\sqrt{\alpha _r^2+\mu _r^2}} \\
		\frac{\alpha _r+ i\mu_r}{\sqrt{\alpha _r^2+\mu _r^2}} & 0 \\
	\end{array} \right) . \label{Pnot}
\end{equation}
The eigenvalues of ${\cal P}$ are $\pm 1$ so that it is indeed negative definite. The energy eigenvalues of $H$ 
\begin{equation}
	E_\pm =\frac{1}{2} \left[ - \omega \pm \frac{1}{\alpha_r}\sqrt{\Delta }         \right] \qquad \Delta :=  (\alpha_r^2 + \mu_r^2)(\alpha_r^2 - \mu_i^2)-\alpha_r^2 \tau_i^2 ,
\end{equation}
may become real, complex conjugate or coalesce depending on the value of the discriminant, exhibiting the three possible ${\cal PT}$-regimes in parameter space. These are the $\cal{PT}$-symmetric regime when $\Delta>0$, the exceptional point when $\Delta =0$ and the spontaneously broken $\cal{PT}$-regime when $\Delta<0$.

\subsection{Hermitian Dyson map,  \protect{ $ [ \tilde{{\cal P}}{\cal T} ,H   ] = 0 $ } }
Let us next turn to the fully time-dependent case and solve the time-dependent Dyson equation (\ref{TDDE}) for $\eta(t)$ and $h(t)$. Substituting the Hermitian Ansatz 
\begin{equation}
	\eta(t)= \eta_0(t)  \mathbb{I} + \eta_z(t) \sigma_z, \qquad  \eta_0,\eta_z \in \mathbb{R}
\end{equation}
into the (\ref{TDDE}) and demanding the left hand side to be Hermitian, leads to the two coupled first order differential equations with an additional constraints (\ref{const1}) and
\begin{equation}
	\dot{\eta}_0= \frac{1}{2} \eta_z \tau_i, \qquad \dot{\eta}_z= \frac{1}{2} \eta_0 \tau_i, \qquad \mu_i = - \frac{2 \alpha_r \eta_0 \eta_z}{\eta_0^2 + \eta_z^2} . \label{cdiff}
\end{equation}
The first two equations can be combined into the second order equation
\begin{equation}
	\ddot{\eta}_0 -\frac{\dot{\tau}_i}{\tau_i} \dot{\eta}_0- \frac{1}{4} \tau_i^2 \eta_0=0,
\end{equation}
which is solved by
\begin{equation}
	\eta_0(t) = c_1 \sinh\left[ \frac{\delta(t)}{2} \right] + c_2 \cosh\left[ \frac{\delta(t)}{2} \right], \qquad \text{with} \quad \delta(t):= \int^t \tau_i(s)ds,
\end{equation}
and real integration constants $c_1$ and $c_2$. It then follows directly from the constraints (\ref{cdiff}) that $	\eta_z(t) = c_1 \cosh\left[ \delta(t)/2 \right] + c_2 \sinh\left[ \delta(t)/2 \right]$ and therefore the time-dependent Dyson map becomes
\begin{equation}
	\eta(t)=\left(
	\begin{array}{cc}
		(c_1+ c_2) \exp\left[ \frac{\delta(t)}{2}   \right] & 0 \\
	0     & (c_1- c_2) \exp\left[-\frac{\delta(t)}{2}   \right]   \\
	\end{array}
	\right).
	\label{solDyson}
\end{equation}
Since $	\eta(t)$ is Hermitian the corresponding metric operator is immediately obtained as
\begin{equation}
	\rho(t)=\eta^2(t)=\left(
	\begin{array}{cc}
		(c_1+ c_2)^2 e^{\delta(t)} & 0 \\
		0     & (c_1- c_2)^2e^{-\delta(t)}   \\
	\end{array}
	\right),
	\label{solmetric}
\end{equation}
which is obviously positive definite with $\det\rho = (c_1^2- c_2^2)^2 $.

 The Hermitian Hamiltonian results from (\ref{TDDE}) to  
\begin{equation}
	h(t)=\left(
	\begin{array}{cc}
		- \frac{\omega}{2} & \frac{\left(c_1^2-c_2^2\right) \left[\alpha _r(t)-i \mu _r(t)\right]}{4 c_1 c_2 \sinh [\delta (t)]+2
			\left(c_1^2+c_2^2\right) \cosh [\delta (t)]} \\
		\frac{\left(c_1^2-c_2^2\right) \left[\alpha _r(t)+i \mu _r(t)\right]}{4 c_1 c_2 \sinh [\delta (t)]+2
			\left(c_1^2+c_2^2\right) \cosh[\delta (t)]}      & 	- \frac{\omega}{2}  \\
	\end{array}
	\right).
	\label{hermh}
\end{equation}
The corresponding instantaneous energy eigenvalues together with their normalised eigenstates are
\begin{equation}
\tilde{E}_\pm(t) = \pm\frac{\sqrt{2} \left(c_1^2-c_2^2\right) \sqrt{\alpha _r^2(t) +\mu _r^2(t)}}{4 c_1 c_2 \sinh [\delta  (t)]+2 \left(c_1^2+c_2^2\right)
	\cosh [\delta(t)  ]}-\frac{\omega }{2}, \quad \vert \tilde{\chi}_\pm(t) \rangle = \frac{1}{\sqrt{2}}\left(
\begin{array}{c}
\frac{\sqrt{\alpha _r^2(t)+\mu _r^2(t)}}{\alpha _r(t)+  i \mu _r(t)} \\
1 \\
\end{array}
\right).
	\label{insten}
\end{equation}
With these expressions we can now directly compute the time-dependent $\tilde{{\cal P}}$, $\tilde{{\cal C}}$ operators and the Berry phase (\ref{BerrynH1}). From the definition (\ref{defct}), together with $\vert \tilde{\phi}_\pm(t) \rangle  = \eta^{-1}(t) \vert \tilde{\chi}_\pm(t) \rangle  $ and the signature $s_{\pm} = \pm 1$ we compute
\begin{equation}
  \tilde{{\cal C}}(t) = \left(
	\begin{array}{cc}
		0 & \frac{\left(c_1-c_2\right) e^{-\delta(t) } \sqrt{\alpha _r^2(t)+\mu _r^2(t)}}{\left(c_1+c_2\right) \left[\alpha _r(t)+i \mu_r(t)\right]} \\
		\frac{\left(c_1+c_2\right) \left[\alpha _r(t)+i \mu _r(t)\right] e^{\delta(t) } }{\left(c_1-c_2\right)
			\sqrt{\alpha _r^2(t)+\mu _r^2(t)}} & 0 \\
	\end{array}
	\right),
\end{equation}
which is an involution operator that commutes with the non-Hermitian energy operator
\begin{equation}
	\tilde{H}(t)=\left(
	\begin{array}{cc}
		-\frac{\omega }{2} & -\frac{\left(c_1-c_2\right)^2 \left[\alpha _r(t)-i \mu _r(t)\right]}{\left(c_1+c_2\right)^2
			e^{2 \delta (t)}+\left(c_1-c_2\right){}^2} \\
		-\frac{\left(c_1+c_2\right){}^2 e^{\delta (t)} \left[\alpha _r(t)+i \mu _r(t)\right]}{4 c_2 c_1 \sinh \left[\delta (t) \right]+2 (
			c_1^2 + c_2^2 ) \cosh \left[\delta (t)\right]} & -\frac{\omega }{2} \\
	\end{array}
	\right)
\end{equation}
 as stated in (\ref{cprop}). Next we calculate the time-dependent operator $\tilde{{\cal P}}(t)$ from relation (\ref{defct}) to
\begin{equation}
\tilde{{\cal P}}(t) =\rho(t) \tilde{{\cal C}}(t) =\frac{c_1^2-c_2^2}{\sqrt{\alpha _r^2(t)+\mu _r^2(t)}}  \left(
\begin{array}{cc}
	0 & \alpha _r(t)-i \mu _r(t) \\\alpha _r(t)+i \mu _r(t) & 0 \\
\end{array}
\right).
\end{equation}
We notice that $\tilde{{\cal P}}^2(t) \neq \mathbb{I} $, although this can be achieved with a particular choice for the constants, e.g. $c_1=1, c_2=0$. We also verify that $\tilde{{\cal P}}(t)$ does indeed satisfy the second version of the time-dependent quasi-Hermiticity relation in (\ref{TDQHE}). However, with eigenvalues $(c_1^2-c_2^2)$ and $-(c_1^2-c_2^2)$, it is not positive and therefore not a metric. We verify that the non-Hermitian time-dependent energy operator is i) quasi-Hermitian with regard to the action of the intertwining operator $\tilde{{\cal P}}(t)$, ii) the $\tilde{{\cal P}}$-operator converts right eigenstates of $\tilde{H}$ into their left eigenstates as $\tilde{{\cal P}}(t)  \vert \tilde{\psi}_\pm \rangle  = \pm   \vert \tilde{\phi}_\pm \rangle$, with the bi-orthonormality preserved when scaling the states as $\vert \tilde{\psi} \rangle \rightarrow 1/\alpha \vert \tilde{\psi} \rangle$, $\vert \tilde{\phi} \rangle \rightarrow \alpha \vert \tilde{\phi} \rangle$ , and iii) that $\tilde{{\cal P}}(t)$ is Hermitian. This means all three relations in (\ref{PTrel}) hold and hence the instantaneous energy eigenvalues of the non-Hermitian energy operator are guaranteed to be real.

In addition we notice that each term in the energy operator is $\tilde{{\cal P}}(t)$-quasi Hermitian, so that relations $i)$ and $iii)$ also hold for the Hamiltonian $H(t)$. However, the second relation in (\ref{PTrel}) does not hold for $H(t)$ as 
\begin{equation}
	 {\cal P}(t)  \vert \hat{\psi}_\pm \rangle  \neq \alpha \vert \hat{\phi}_\pm \rangle,
\end{equation}
so that we can not apply the above argument to the Hamiltonian.
Thus in this sense the broken ${\cal P}{\cal T}$-regime in the time-independent scenario has been mended in the time-dependent case, a possibility first observed in \cite{AndTom3}.

Having computed the eigenstates of $\tilde{H}$ and the Dyson map $\eta(t)$ we may now also directly calculate the Berry phase by means of (\ref{BerrynH1}). We find 
\begin{equation}
	\gamma_\pm = \frac{1}{2} \int_0^T   \frac{\alpha_r \dot{\mu_r}- \mu_r \dot{\alpha}_r}{\alpha _r^2 +\mu _r^2} dt = \left. \frac{1}{2} \arctan\left( \frac{\mu_r}{\alpha_r}\right) \right\vert_0^T . \label{Berryph2}
\end{equation}
Evidently $\gamma_\pm$ is always real.

\subsection{Non-Hermitian Dyson map, \protect{ $ [ \tilde{{\cal P}}{\cal T} ,H   ] \neq 0 $ } }

In the previous section we have constructed a relatively simple diagonal solution for a Hermitian Dyson map $\eta(t)$. We also found that $ [ \tilde{{\cal P}}{\cal T} ,H   ] = 0 $ and for a particular choice of the constants $\tilde{{\cal P}}(t)$-operator became identical to the parity operator in the time-independent case. However, it is well-known that Dyson maps are not unique in general and one may even construct infinite series in the time-dependent case \cite{fring2021infinit}. We will now exploit this ambiguity and see if one can construct $\tilde{{\cal P}}(t)$-operator that are more distinct from the parity operator by making a more generic non-Hermitian Ansatz and solve (\ref{TDDE}) once more. We assume now 
\begin{equation}
	\eta(t)= \eta_0(t)  \mathbb{I} + \eta_z(t) \sigma_z + i \eta_y(t) \sigma_y, \qquad  \eta_0,\eta_y,\eta_z \in \mathbb{R},
\end{equation}
and proceed as in the previous example to solve the time-dependent Dyson equation. In this case we find the solution 
\begin{equation}
	\eta(t)= \left(
	\begin{array}{cc}
		\frac{c_1 \mu _i}{\alpha _r}-2 c_1 & \frac{c_1 \mu _i}{\alpha _r} \\
		-\frac{c_1 \mu _i}{\alpha _r} & -\frac{c_1 \mu _i}{\alpha _r}-2 c_1 \\
	\end{array}
	\right)= -2 c_1  \mathbb{I} + \frac{c_1 \mu_i}{\alpha_r} \sigma_z + i \frac{c_1 \mu_i}{\alpha_r}\sigma_y, \label{neweta}
\end{equation}
with $c_1$ being an integration constant and the additional constraints
\begin{equation}
\mu_r=-\tau _i -2 A, \qquad  \alpha_i = 2 \frac{\mu_i}{\alpha_r} A,
\qquad \tau_r = \mu_i, \qquad A:=\frac{\tau _i \alpha
	_r^2}{\mu _i^2}-\frac{\dot{\alpha}_r}{\mu _i}+\frac{\alpha _r \dot{\mu}_i}{\mu _i^2} ,
\end{equation}
together with the time-dependent Hermitian Hamiltonian 
\begin{equation}
	h = \left(
	\begin{array}{cc}
		-\frac{\omega }{2} & -\frac{\alpha _r}{2}-i A \\
		-\frac{\alpha _r}{2}+i A & -\frac{\omega }{2} \\
	\end{array}
	\right).
\end{equation}
We have omitted here the calculation, but the critical reader may convince themselves by simply substituting the solution back into the time-dependent Dyson equation. Notice also that the solution (\ref{neweta}) does not reduce to the previous one, as the elimination of the non-Hermitian term by $\mu_i \rightarrow 0$ will simply lead to the trivial solution proportional to $\mathbb{I}$.

The time-dependent instantaneous energy eigenvalues together with their normalised eigenstates as introduced in (\ref{12inst}) are found to
\begin{equation}
	\tilde{E}_\pm = \frac{1}{2} \left( -\omega \pm \sqrt{4 A^2+\alpha _r^2} \right), \qquad  \vert \tilde{\chi}_\pm \rangle = \frac{1}{\sqrt{2}} \left(
	\begin{array}{c}
		\frac{\mp i \sqrt{4 A^2+\alpha _r^2}}{ 2 A+i \alpha _r } \\
		1 \\
	\end{array}
	\right) .
\end{equation}
Evidently $\tilde{E}_\pm  \in \mathbb{R}$, such that the broken regime from the time-independent regime has been mended once more. 
As in the previous section we use the relation $\vert \tilde{\psi}_\pm(t) \rangle  = \eta^{-1}(t) \vert \tilde{\chi}_\pm(t) \rangle  $ between the eigenstates of $h(t)$ and $\tilde{H}(t)$ to compute the states in the expansion (\ref{defc}). For the signature $s_{\pm} = \pm 1$ we obtain 
\begin{equation}
	\tilde{{\cal C}} = \frac{-1}{\sqrt{4 A^2+\alpha _r^2}} \left(
	\begin{array}{cc}
		\mu _i+ i\frac{  \mu _i^2}{\alpha _r^2}A & \alpha _r+\mu _i+i \left(\frac{ \mu _i^2}{\alpha _r^2}+\frac{2  \mu _i}{\alpha _r}+2
		\right)A \\
		\alpha _r-\mu _i-i \left(\frac{ \mu _i^2}{\alpha _r^2}-\frac{2  \mu _i}{\alpha _r}+2 \right)A & -\mu _i- i\frac{  \mu _i^2}{\alpha _r^2}A \\
	\end{array}
	\right).
\end{equation}
We verify that this operator squares to $\mathbb{I}$ and does indeed commute with the energy operator
\begin{equation}
	\tilde{H}= \left(
	\begin{array}{cc}
		-\frac{\mu _i}{2}-\frac{\omega }{2} -i\frac{ A \mu _i^2}{2 \alpha _r^2} &-\frac{\mu _i}{2}-\frac{\alpha _r}{2}  -i A \left(\frac{\mu _i^2}{2 \alpha
			_r^2}+\frac{\mu _i}{\alpha _r}+1\right) \\
		\frac{\mu _i}{2}-\frac{\alpha _r}{2}+ i A \left(\frac{\mu _i^2}{2 \alpha _r^2}-\frac{\mu _i}{\alpha _r}+1\right) &
		\frac{\mu _i}{2}-\frac{\omega }{2} + i \frac{ A \mu _i^2}{2 \alpha _r^2}\\
	\end{array}
	\right),
\end{equation}
as required in (\ref{cprop}).

The corresponding metric is directly computed to
\begin{equation}
	\rho = \eta^\dagger \eta =\left(
	\begin{array}{cc}
		\frac{2 c_1^2 \left(\mu _i^2-2 \mu _i \alpha _r+2 \alpha _r^2\right)}{\alpha _r^2} & \frac{2 c_1^2 \mu _i^2}{\alpha
			_r^2} \\
		\frac{2 c_1^2 \mu _i^2}{\alpha _r^2} & \frac{2 c_1^2 \left(\mu _i^2+2 \mu _i \alpha _r+2 \alpha _r^2\right)}{\alpha
			_r^2} \\
	\end{array}
	\right) .
\end{equation}
The eigenvalues of $\rho$ are
\begin{equation}
\lambda_\pm^\rho = \frac{2 c_1^2 \left[\mu _i^2 \alpha _r^4 +2 \alpha
	_r^6 \pm\sqrt{\mu _i^2 \alpha _r^8 \left(\mu _i^2+4 \alpha _r^2\right)}\right]}{\alpha _r^6} \geq 0,
\end{equation}
so that the metric is indeed positive.

We can now compute the time-dependent parity operator according to (\ref{defct}) as
\begin{equation}
 \tilde{{\cal P}}(t)= \rho(t)\tilde{{\cal C}}(t) = \frac{2c_1^2}{ \alpha _r \sqrt{4 A^2+\alpha _r^2}} \left(
 \begin{array}{cc}
 	\mu _i \left(\mu _i-2 \alpha _r\right) & \mu _i^2-2 \alpha _r^2-4 i A \alpha _r \\
 	\mu _i^2-2 \alpha _r^2+4 i A \alpha _r & \mu _i \left( \mu _i+2\alpha _r\right) \\
 \end{array}
 \right).
\end{equation}
The two eigenvalues of $\tilde{{\cal P}}$ are 
\begin{equation}
     \lambda_\pm^{{\cal P}} = \frac{2c_1^2}{ \alpha _r \sqrt{ 4 A^2+\alpha _r^2}} \left(\mu_i^2 \pm \sqrt{16 A^2 \alpha_r^2 + 4 \alpha_r^4 + \mu_i^4 }    \right),
\end{equation}
so that $\tilde{{\cal P}}$ is negative definite with $\det \tilde{{\cal P}} =-16 c_1^4 $. Moreover, we verify that the  first relation in (\ref{PTrel}) is satisfied, that $ \tilde{{\cal P}} \vert \tilde{\psi}_\pm \rangle = \pm \vert \tilde{\phi}_\pm \rangle $ and that $\tilde{{\cal P}} =\tilde{{\cal P}}^\dagger $. These properties suffice to guarantee the reality of the instantaneous energy eigenvalues $\tilde{E}_\pm$.

The Berry phase is computed in a similar way as in the previous section from (\ref{BerrynH1}) to 
\begin{equation}
	\gamma_\pm = -\frac{1}{2} \int_0^T   \frac{ 2 \alpha_r \dot{A} - 2A \dot{\alpha}_r}{4 A^2 + \alpha_r^2} dt =
	 \left. -\frac{1}{2} \arctan\left( \frac{2 A}{\alpha_r}\right) \right\vert_0^T , \label{Berryph3}
\end{equation}
which is always real.

\section{Conclusions}
We have identified a new operator $\tilde{{\cal P}}$ as the non-involutory non-positive definite solution of the quasi-Hermiticity relation (\ref{TDQHE}) involving the energy operator. The existence of this operator will ensure the reality of the instantaneous energies when it satisfies the three properties stated in (\ref{PTrel}). As this operator differs in general from the operator ${\cal P} {\cal T}$-operator of the time-independent case, the characteristic regimes this operator classifies are naturally different from those in the time-dependent case.

Furthermore, we showed that when changing the commonly used adiabatic approximation from expanding the solutions to the TDSE in terms of the eigenstates of the energy operator rather than the Hamiltonian, the Berry phase will be identical to the one computed for the Hermitian counterpart and therefore real.

\medskip
\noindent \textbf{Acknowledgments:} RT is supported by EPSRC grant EP/W522351/1.  TT is supported by JSPS KAKENHI Grant Number JP22J01230.

\newif\ifabfull\abfulltrue


\end{document}